\documentclass{JHEP3}
\usepackage{graphicx}
\usepackage{amsmath}\usepackage{amssymb}
\usepackage{epstopdf}

\makeatletter
\newdimen\rh@wd
\newdimen\rh@hta
\newdimen\rh@htb
\newbox\rh@box
\def\rh@measure#1{\setbox\rh@box=\hbox{$#1$}\rh@wd=\wd\rh@box
\rh@hta=\ht\rh@box}

\def\widecheck#1{\rh@measure{#1}%
   \setbox\rh@box=\hbox{$\widehat{\vrule height \rh@hta width\z@
\kern\rh@wd}$}%
   \rh@htb=\ht\rh@box \advance\rh@htb\rh@hta \advance\rh@htb\p@
   \ooalign{$\vrule height \ht\rh@box width\z@ #1$\cr
            \raise\rh@htb\hbox{\scalebox{1}[-1]{\box\rh@box}}\cr}}
\makeatother
\makeatletter
\def\widebar{$\m@th\mathord=\mkern-7mu
  \cleaders\hbox{$\!\mathord=\!$}\hfill
  \mkern-7mu\mathord=$}
\makeatother

\title{Symmetries of the Self-Dual Sector of N=4 Super Yang-Mills on the Light Cone} 
\author{Adam Wardlow  \\
Department of Mathematical Sciences, University of Durham\\
South Road, Durham, DH1 3LE, U.K.\\ 
E-mail: \email{a.b.wardlow@durham.ac.uk}
}
\abstract{A recent paper proposes a way of constructing infinite dimensional symmetries of the non-supersymmetric self-dual Yang-Mills action using isometries of the space-time. We review the Lagrangian formulation of $N=4$ super Yang-Mills MHV rules and extend the approach taken for the non-supersymmetric case to construct infinite dimensional symmetries of self-dual $N=4$ super Yang-Mills.  }
\preprint{DCPT-09/67}
\begin{document}
\numberwithin{equation}{section}
\section{Introduction}
The normal Feynman approach to calculating $n$ gluon tree-level  scattering amplitudes is well understood but the complexity of calculations grows quickly with $n$ making the method inefficient and prohibitive. It was recently observed that such gluon amplitudes localise on simple curves in twistor space \cite{witten-2004-252} and this led to a new set of rules for calculating them \cite{cachazo}. This approach provides an  alternative to the Feynman rules with drastically reduced complexity, for example the Parke-Taylor amplitude for tree level scattering of $n-2$ positive helicity gluons and $2$ negative is remarkably simple \cite{Parke:1986gb}. The new set of rules was initially proven outside the Lagrangian formalism using the BCFW recursion relation \cite{britto-2005-94} and using twistor methods, (See \cite{boels-2007-014} through to \nocite{mason-2006-636,boels-2007-648,boels-2007-76,boels-2007-016,boels-2008-007,boels-2008-183} \cite{Wen}). More recently they have been derived in the non-supersymmetric theory by applying a non-local canonical transformation to the Yang-Mills action on the light cone \cite{Gorsky:2005sf, Pauls_paper}. The action is split into the Chalmers-Siegel action describing the self-dual sector \cite{Chalmers:1996rq} plus the rest and the canonical transformation maps the self-dual part of the action to a free action. The transformation was also studied in more detail in \cite{ettle_and_morris}.

This Lagrangian approach to the derivation of the MHV rules was extended to $N=4$ supersymmetric Yang-Mills theory in 4 dimensions in \cite{huang} where they write the N=4 SYM action derived in \cite{brink} in terms of superfields $\Phi$ and $\bar{\Phi}$. Using the CPT self conjugacy property of the superfields they write down the classically free, self-dual part of the action. They then extend the canonical transformation of \cite{Pauls_paper} to map this to a free theory. The rest of the action gives us the interacting terms.

In a recent paper \cite{me} the transformation and its inverse is used to construct infinite dimensional non-local symmetries of the self-dual part of the non-supersymmeteric Yang-Mills theory off-shell. (Also see \cite{dolan} where Dolan constructs symmetries on-shell and \cite{adam}, \cite{Popov:2006qu} and \cite{Wolf:2004hp}).  They use the simple fact that a free theory with Euler-Lagrange equation $\Omega(x)\phi(x)=0$ has a symmetry if $\Omega(x^G)=\Omega(x)$ where $x\rightarrow x^G$ is a finite isometry of the space-time. Then since $0=\Omega(x^G)\phi(x^G)=\Omega(x)\phi(x^G)$, we see that $\phi(x^G)$ is a new solution. Because of the linearity of the free Euler-Lagrange equation we can construct a new solution as $\phi(x)+\epsilon\phi(x^G)$. This leads to higher order conserved currents such as the Zilch of the electromagnetic field \cite{lipkin} and those calculated in \cite{fairlie}. More generally, we could consider the case where $x\rightarrow x^G$ is a conformal tranformation with $f(x)\Omega(x^G)=\Omega(x)$. In the paper \cite{popov-1996-374} the authors construct infinite-dimensional symmetries of the self-dual Yang-Mills equations based on conformal symmetries, we however shall concentrate on isometries.

In this paper we shall extend this and construct symmetries of the $N=4$ self-dual SYM action by using the canonical transformation to map the self-dual action to the free theory and then writing the symmetry in terms of the free fields. We derive an expression for the inverse transformation and use it to write the expression in terms of the original variables. We examine the first 4 orders in powers of the fields and then hypothesise the general result. We then prove the above expression leaves the action invariant and conclude by showing how we can extract expressions for the transformations of the component fields.
\section{Review of $N=4$ Super Yang-Mills on the Light Cone and the MHV rules Lagrangian}
\subsection{Light Cone N=4 SYM}
We shall review the construction of the $N=4$ supersymmetric Yang-Mills action on the light cone. For a more detailed treatment see \cite{brink, huang}. Let us start by considering the action in 10 dimensions, which is given by
\begin{equation}
\label{eq:10d susy action}
S=\int d^{10}x \left\{\frac{1}{4}F_{a}^{\mu\nu}F_{\mu\nu a}+\frac{1}{2}i\bar{\psi}^{a}\Gamma^{\mu}D_{\mu}\psi^a\right\}
\end{equation}
for $\mu,\nu=0,\cdots,9$ and where $\Gamma$ is a generalisation of the Dirac gamma matrices to 10 dimensions. The spinor degrees of freedom satisfy the Weyl and Majorana conditions and $F_{\mu\nu}^a$ is given by
\begin{equation}
\nonumber
F_{\mu\nu}^a=\partial_{\mu}A_{\nu}^{a}-\partial_{\nu}A_{\mu}^{a}+gf^{abc}A_{\mu}^b A_{\nu}^c.
\end{equation}
As stated in \cite{brink} it is straightforward to show that the action (\ref{eq:10d susy action}) is invariant under the supersymmetry transformations
\begin{equation}
\nonumber
\delta A_{\mu}=\bar{\xi}\Gamma_{\mu}\psi \ \ \ \ \ \ \ \ \ \ \ \ \ \ \ \ \ \ \delta\psi=-\frac{1}{2}F_{\mu\nu}\Gamma^{\mu\nu}\xi.
\end{equation}
It is known however that consecutive supersymmetry transformations of this form do not close to form an algebra off-shell. To make this algebra close requires the introduction of auxiliary fields, however as explained in \cite{brink} it is not known how to do this. Take for example the commutator of transformations of the spinor field 

\begin{equation}\begin{split}
\label{eq:susy comm}
\left(\delta_{\xi_2}\delta_{\xi_1}-\delta_{\xi_1}\delta_{\xi_2}\right)\psi=\left(\bar{\xi}_2\Gamma^{\mu}\xi_1\right)D_{\mu}\psi-\frac{1}{2}\left(\bar{\xi}_2\Gamma^{\mu}\xi_1\right)\Gamma_{\mu}\Gamma^{\nu}D_{\nu}\psi
\end{split}\end{equation}
which, by using the field equation $\Gamma^{\mu}D_{\mu}\psi=0$, closes to
\begin{equation}\begin{split}
\nonumber
\left(\delta_{\xi_2}\delta_{\xi_1}-\delta_{\xi_1}\delta_{\xi_2}\right)\psi=\left(\bar{\xi}_2\Gamma^{\mu}\xi_1\right)D_{\mu}\psi.
\end{split}\end{equation}
It is still possible to retain half the susy on-shell at this stage by transforming to a frame in which only $\hat{p}$ is non vanishing.  As explained in the papers \cite{brink, huang}, if we now split the spinor as follows
\begin{equation}
\nonumber
\psi=-\frac{1}{2}\left(\widehat{\Gamma}\widecheck{\Gamma}+\widecheck{\Gamma}\widehat{\Gamma}\right)\psi=\widehat{\psi}+\widecheck{\psi}
\end{equation}
where $\hat{\Gamma}={1}/{\sqrt{2}}\left(\Gamma^0+\Gamma^1\right)$ and $\check{\Gamma}={1}/{\sqrt{2}}\left(\Gamma^0-\Gamma^1\right)$ then (\ref{eq:susy comm}) now closes with on-shell degrees of freedom $A_{\perp}$ and $\widecheck{\psi}$ leaving only the $SO(8)$ subgroup of the original Lorentz group manifest \cite{huang}. Now (L.Brink \textit{et al}) \cite{brink} dimensionally reduce this to four dimensions which breaks the $SO(8)$ invariance
\begin{equation}
SO(8)\rightarrow SO(6)\otimes SO(2)\sim SU(4)\otimes U(1)
\end{equation} leaving the 4 dimensional SUSY algebra
\begin{equation}
\label{eq:susy algebra}
\left\{\bar{q}^A,q_{B}\right\}=-i\sqrt{2}\delta^{A}_{B}\hat{\partial}
\end{equation}
where $A$ and $B$ are $SU(4)$ indices $A,B=1,2,3,4$. A supersymmetry transformation on superspace $\left(\hat{x},\check{x},\tilde{x},\bar{x} ;\theta,\bar{\theta}\right)$ generates the following change in coordinates
\begin{equation}
\nonumber
 \left(\hat{x},\check{x},\tilde{x},\bar{x}; \theta,\bar{\theta}\right)\rightarrow\left(\hat{x}+\frac{i}{\sqrt{2}}\xi^A\bar{\theta}_A-\frac{i}{\sqrt{2}}\theta^{A}\bar{\xi}_A,\check{x},\tilde{x},\bar{x}; \theta+\xi,\bar{\theta}-\bar{\xi}\right)
\end{equation}
where $\theta$ are Grassman variables. The transformations give rise to the following SUSY generators and covariant derivatives, $d$ and $\bar{d}$
\begin{align}
\label{eq:susy gen}
\nonumber q_A&=\frac{\partial}{\partial\theta^A}+\frac{i}{\sqrt{2}}\bar{\theta}_A\hat{\partial} & \bar{q}^A&=-\frac{\partial}{\partial\bar{\theta}_A}-\frac{i}{\sqrt{2}}\theta^A\hat{\partial}\\
d_A&=\frac{\partial}{\partial\theta^A}-\frac{i}{\sqrt{2}}\bar{\theta}_A\hat{\partial} & \bar{d}^A&=-\frac{\partial}{\partial\bar{\theta}_A}+\frac{i}{\sqrt{2}}\theta^A\hat{\partial}
\end{align}
and it is easily verified that $q$ and $\bar{q}$ do indeed satisfy the SUSY algebra given in \cite{brink} and in (\ref{eq:susy algebra}). A chiral superfield is defined by imposing the constraint
\begin{equation}
\label{eq:constraint1}
\bar{d}_A\Phi=0
\end{equation}
and further, the N=4 susy multiplet is CPT self conjugate and so we impose a second `reality' constraint in the same way that was discussed in \cite{brink}, 
\begin{equation}
\label{eq:constraint2}
\bar{\Phi}=\frac{\epsilon^{ABCD}}{48\hat{\partial}^2}d_A d_B d_C d_D\Phi.
\end{equation}
A superfield satisfying both (\ref{eq:constraint1}) and (\ref{eq:constraint2}) is written
\begin{equation}\begin{split}
\label{eq:sfield}
\Phi(x,\theta,\bar{\theta})=&\frac{1}{\hat{\partial}}A(y)+\frac{i}{\hat{\partial}}\theta^A\lambda_A(y)+i\frac{1}{\sqrt{2}}\theta^A\theta^B \bar{C}_{AB}(y)\\&+\frac{\sqrt{2}}{3!}\theta^A\theta^B\theta^C\epsilon_{ABCD}\bar{\lambda}^D(y)+\frac{1}{12}\theta^A\theta^B\theta^C\theta^D\epsilon_{ABCD}\hat{\partial}\bar{A}(y)
\end{split}\end{equation}
where $y=\left(\hat{x}-\frac{i}{\sqrt{2}}\theta^A\bar{\theta}_A,\check{x},\tilde{x},\bar{x}\right)$ is known as the chiral basis in which (\ref{eq:constraint1}) is trivially satisfied and the fields $A,\lambda$ and $C$ are the gauge fields, fermions and scalars respectively. (See \cite{bagger}, page (30)). In terms of this superfield the N=4 super Yang-Mills action on the light cone in 4 dimensions is
\begin{equation}\begin{split}
\label{eq:N=4 superfield SYM action}
S=tr\int d^4 x d^4\theta d^4\bar{\theta}\bigg\{&\bar{\Phi}\frac{\hat{\partial}\check{\partial}-\tilde{\partial}\bar{\partial}}{\hat{\partial}^2}\Phi+\frac{2}{3}gf^{abc}\left[\frac{1}{\hat{\partial}}\bar{\Phi}^a\Phi^b\bar{\partial}\Phi^c+complex\ conjugate\right]\\
&-\frac{g^2}{2}f^{abc}f^{ade}\left[\frac{1}{\hat{\partial}}\left(\Phi^b\hat{\partial}\Phi^c\right)\frac{1}{\hat{\partial}}\left(\bar{\Phi}^d\hat{\partial}\bar{\Phi}^e\right)+\frac{1}{2}\Phi^b\bar{\Phi}^c\Phi^d\bar{\Phi}^e\right]\bigg\}
\end{split}\end{equation}
as given in \cite{brink} and \cite{huang}. It is straightforward to express this in component form which agrees with the expression in \cite{brink}, (Equation (3.13) in their paper).

\subsection{MHV Rules Lagrangian for N=4 SYM}
\label{sec:mhvruled}
Let us examine the helicity content of the action by considering each part. We write
\begin{equation}
\nonumber
S=S^{-+}+S^{-++}+S^{--+}+S^{--++}
\end{equation}
with
\begin{align}
\nonumber S^{-+}&=tr\int d^4 x d^4\theta d^4\bar{\theta}\big\{\bar{\Phi}\frac{\hat{\partial}\check{\partial}-\tilde{\partial}\bar{\partial}}{\hat{\partial}^2}\Phi\big\}\\
\nonumber S^{-++}&=tr\int d^4 x d^4\theta d^4\bar{\theta}\big\{\frac{2}{3}gf^{abc}\frac{1}{\hat{\partial}}\bar{\Phi}^a\Phi^b\bar{\partial}\Phi^c\big\}
\end{align}
and so on for $S^{--+}$ and $S^{--++}$. In the MHV rules (Maximal helicity violating amplitude) an n point amplitude consists of 2 negative helicities and n-2 positive helicities (see \cite{cachazo, Parke:1986gb, boels-2007-648}). In parallel with papers by Mansfield, and Ettle and Morris \cite{Pauls_paper, ettle_and_morris}, the part of the action $S_{-++}$ clearly does not satisfy this requirement and further, terms with more than two positive helicities are missing from the full action (\ref{eq:N=4 superfield SYM action}).

We can express (\ref{eq:N=4 superfield SYM action}) in the chiral basis $y$ by expressing the action in terms of $\Phi$ only using (\ref{eq:constraint2}) at the expense of introducing covariant derivatives in to the action. One will get a kinetic term, a cubic term with 4 covariant derivatives and a further two terms with 8 covariant derivatives, as explained in \cite{huang}. Chalmers and Siegel \cite{Chalmers:1996rq} show that terms which contain only four covariant derivatives, i.e. $S^{-+}+S^{-++}$ express the self-dual sector in terms of the Chalmers-Siegel action. Classically, self-dual Yang-Mills is free so we wish to transform the self-dual sector $S^{-+}+S^{-++}$ into a free action by a canonical change of fields $\Phi[\chi]$. This procedure absorbs the unwanted term $S^{-++}$ into a free action, and it turns out the change of field variables generates all the missing terms $S^{--+\cdots +}$.
By that argument, Feng and Huang give us the Chalmers-Siegel action describing the self-dual sector as
\begin{equation}\begin{split}
\label{eq:Ssd}
S_{SD}&=tr\int d^4 x d^4\theta\big\{\Phi\left(\hat{\partial}\check{\partial}-\tilde{\partial}\bar{\partial}\right)\Phi+\frac{2}{3}\hat{\partial}\Phi\left[\Phi,\bar{\partial}\Phi\right]\big\}\\
&=tr\int d^4 x d^4\theta\big\{\chi\left(\hat{\partial}\check{\partial}-\tilde{\partial}\bar{\partial}\right)\chi\big\}
\end{split}\end{equation}
where the free superfield $\chi$ is written as
\begin{equation}\begin{split}
\label{eq:free_superfield}
\chi(y,\theta)=&\frac{1}{\hat{\partial}}B(y)+\frac{i}{\hat{\partial}}\theta^A\rho_A(y)+i\frac{1}{\sqrt{2}}\theta^A\theta^B \bar{D}_{AB}(y)\\&+\frac{\sqrt{2}}{3!}\theta^A\theta^B\theta^C\epsilon_{ABCD}\bar{\rho}^D(y)+\frac{1}{12}\theta^A\theta^B\theta^C\theta^D\epsilon_{ABCD}\hat{\partial}\bar{B}(y)
\end{split}\end{equation}
in the chiral basis with $B$ and $\bar{B}$ the gauge fields, $\rho$ and $\bar{\rho}$ are fermions and $D_{AB}$ is a four by four anti-symmetric matrix of real scalars (thus having six independent scalar fields).
The field transformation derived in \cite{huang} satisfies the equation
\begin{equation}
\label{eq:funcdiffeq}
tr\int d^4 x d^4\theta\left\{-\Phi\tilde{\partial}\bar{\partial}\Phi+\frac{2}{3}\hat{\partial}\Phi\left[\Phi,\bar{\partial}\Phi\right]\right\}=tr\int d^4 x d^4\theta \left\{-\chi\tilde{\partial}\bar{\partial}\chi\right\}
\end{equation}
arising from (\ref{eq:Ssd}) and the condition that $\Phi$ and $\chi$ have the same $\check{x}$ dependence (See \cite{Pauls_paper}). Further, we apply the additional constraint
\begin{equation}
\label{eq:normal}
tr\int d^4 x d^4\theta \Phi\hat{\partial}\check{\partial}\Phi=tr\int d^4 x d^4\theta \chi\hat{\partial}\check{\partial}\chi
\end{equation}
as discussed in Feng and Huang, \cite{huang}. They calculate the transformation $\Phi[\chi]$ in their paper but not the inverse transformation which we shall also need. We shall state their result here and calculate the inverse for ourselves using their procedure. Their field redefinition reads
\begin{equation}
\label{eq:phi(chi)}
\Phi_{1}=\chi_{1}+\sum_{n=3}^{\infty}\int_{2\cdots n}C(12\cdots n)\chi_{\bar{2}}\chi_{\bar{3}}\cdots\chi_{\bar{n}}\
\end{equation}
where we use the abbreviations $\Phi_{i}=\Phi(p_i)$ and $\Phi_{\bar{i}}=\Phi(-p_i)$ as we shall do throughout this paper and we drop the momentum conserving delta function $\delta\left(p_1+p_2+\cdots+p_n\right)$. In the above we use the notation
\begin{equation}
\nonumber	
\int_{1 \cdots n}=\int \frac{d^4 p_1}{(2\pi)^4} \cdots \frac{d^4 p_n}{(2\pi)^4}.
\end{equation}
The kernel $C$ is given by
\begin{equation}
\label{eq:C}
C(12\cdots n)=(-1)^n\frac{\hat{2}\hat{3}^2\hat{4}^2\cdots\widehat{n-2}^2\widehat{n-1}^2\hat{n}}{\left(2,3\right)\left(3,4\right)\cdots\left(n-1,n\right)}
\end{equation}
where the bracket $(\ ,\ )$ is given by $(i,j)=\hat{i}\tilde{j}-\tilde{i}\hat{j}$. Now let us calculate the inverse field redefinition $\chi[\Phi]$ for ourselves. We guess the form of the expansion as
\begin{equation}
\label{eq:chi(phi)}
\chi_{1}=\Phi_{1}+\sum_{n=3}^{\infty}\int_{2\cdots n}D\left(12\cdots n\right)\Phi_{\bar{2}}\Phi_{\bar{3}}\cdots\Phi_{\bar{n}}.
\end{equation}
Under the field redefinition and the product of superfields, the $A$ fields do not mix with any of the other fields in the multiplet as they are zeroth order in the expansion of $\theta$ in the superfield. We can simply read off the field transformation for the $A$ and $B$ fields.
\begin{equation}
\nonumber
\frac{B_{1}}{i\hat{p}_1}=\frac{A_{1}}{i\hat{p}_1}-\sum_{n=3}^{\infty}\int_{2\cdots n}D(12\cdots n)\frac{A_{\bar{2}}}{i\hat{p}_2}\frac{A_{\bar{3}}}{i\hat{p}_3}\cdots\frac{A_{\bar{3}}}{i\hat{p}_n}.
\end{equation}
We will compare this to the transformation that is in the literature, namely the papers \cite{Pauls_paper}, \cite{ettle_and_morris} and \cite{me}. We have
\begin{equation}
\nonumber
B_{1}=A_{1}-\sum_{n=3}^{\infty}\int_{2\cdots n}(i)^n\frac{\hat{1}}{(1,2)}\frac{\hat{1}}{(1,2+3)}\cdots\frac{\hat{1}}{(1,2+\cdots +(n-1))}A_{\bar{2}}A_{\bar{3}}\cdots A_{\bar{n}}
\end{equation}
giving our expression for $D(12\cdots n)$ as the following
\begin{equation}
\label{eq:D}
D(12\cdots n)=-(-1)^n\frac{\hat{1}^{n-3}\hat{2}\hat{3}\cdots\hat{n}}{\left(1,2\right)\left(1,2+3\right)\cdots\left(1,2+3+\cdots+(n-1)\right)}
\end{equation}
We can prove this expression by substituting it into (\ref{eq:funcdiffeq}), writing down a recursion relation for the coefficients $D$ and showing they satisfy this recursion relation as follows. Given the canonical transformation condition proved in \cite{huang}, namely
\begin{equation}
\nonumber
 \hat{\partial}\Phi=\int d^4 y \frac{\delta\chi(y)}{\delta\Phi(x)}\hat{\partial}\chi(y),
\end{equation}
we substitute this in to (\ref{eq:funcdiffeq}), transform into momentum space and rearrange to arrive at a relation between fields $\chi$ and $\Phi$ given by
\begin{equation}\begin{split}
\nonumber
 \int_{p_1}\omega(p_1)\Phi(p_1)\frac{\delta\chi(p)}{\delta\Phi(p_1)}+\int_{p_1 p_2 p_3}\frac{\left[\hat{p}_2\Phi(p_2),\bar{p}_3\Phi(p_3)\right]}{\hat{p}_1}\delta(p_1-p_2-p_3)\frac{\delta\chi(p)}{\delta\Phi(p_1)}=\\
=\omega(p)\chi(p)
\end{split}\end{equation}
where $\omega(p)=\bar{p}\tilde{p}/\hat{p}$. Then we take the ansatz for the field redefinition (\ref{eq:chi(phi)}) and proceed by substituting this into the above expression to extract the recurrence relation
\begin{equation}
\nonumber
 D^n(1\cdots n)=\frac{1}{\omega_1+\omega_2+\cdots+\omega_n}\sum_{k=2}^{n-1}\left\{k+1,k\right\}D^{n-1}(1,2,\cdots,k-1,k+(k+1),k+2,\cdots,n)
\end{equation} 
and then substituting (\ref{eq:D}) on the right hand side and taking a factor of $D^n(12\cdots n)$ outside the sum we get
\begin{equation}
\nonumber
 -\frac{D^{n}(1\cdots n)}{\hat{1}(\omega_1+\cdots +\omega_n)}\sum_{k=2}^{\infty}\frac{\left\{k,k+1\right\}}{\hat{k}\widehat{k+1}}\left(1,p_{2,k}\right).
\end{equation}
with $\{i,j\}=\hat{i}\bar{j}-\bar{i}\hat{j}$. The expression under the sum then reduces to $-\hat{1}\left(\omega_1+\cdots+\omega_n\right)$ thus proving the result.

As an aside, we can use the inverse field redefinition to calculate the field redefinition $\bar{B}[A,\bar{A}]$, an expression missing from \cite{Pauls_paper}, \cite{ettle_and_morris} and \cite{me}. We calculate the component of (\ref{eq:chi(phi)}) (which is now proven) that is fourth order in $\theta$ and find
\begin{equation}
\nonumber
 \bar{B}_{\bar{1}}=\sum_{n=2}^{\infty}\sum_{k=2}^n\int_{2\cdots n}\frac{\hat{k}}{\hat{1}}\Theta^k(12\cdots n)A_{\bar{2}}\cdots\bar{A}_{\bar{k}}\cdots A_{\bar{n}}
\end{equation}
where 
\begin{equation}
\nonumber
 \Theta^{k}(1\cdots n)=-\frac{\hat{k}}{\hat{1}}\Gamma(1\cdots n).
\end{equation}
A further point to note about (\ref{eq:chi(phi)}) is that since each term in the expansion is linearly independent, and since $\chi$ is a superfield satisfying the constraints (\ref{eq:constraint1}) and (\ref{eq:constraint2}) then it makes sense that each term in the expansion also satisfies these constraints and is therefore a superfield which has the same form as the free field $\chi$ and the SYM field $\Phi$. So if we write the field redefinition as
\begin{equation}
\nonumber
\chi(1)=\Psi^{0}(1)+\Psi^{1}(1)+\Psi^{2}(1)+\cdots
\end{equation}
and so on, with $\Psi^{0}=\Phi$ and defining
\begin{equation}
\label{eq:Psi}
 \Psi^{n-2}=\int_{2\cdots n}D\left(12\cdots n\right)\Phi_{\bar{2}}\Phi_{\bar{3}}\cdots\Phi_{\bar{n}}
\end{equation}
as the individual terms in the field redefinition, then the $\Psi^{n-2}$ trivially satisfies the constraint (\ref{eq:constraint1}) since $\chi$ can be written in the chiral basis in which it contains no $\bar{\theta}$ and we can write the conjugate of $\Psi$ as the following
\begin{equation}
\nonumber
\bar{\chi}(1)=\bar{\Psi}^{0}(1)+\bar{\Psi}^{1}(1)+\bar{\Psi}^{2}(1)+\cdots
\end{equation}
then since $\chi$ satisfies (\ref{eq:constraint2}) we write
\begin{equation}\begin{split}
\nonumber
\bar{\chi}(1)&=\frac{\epsilon^{ABCD}d_{A}d_{B}d_{C}d_{D}}{48\hat{1}^2}\{\Psi^{0}(1)+\Psi^{1}(1)+\Psi^{2}(1)+\cdots\}\\
&=\frac{\epsilon^{ABCD}d_{A}d_{B}d_{C}d_{D}}{48\hat{1}^2}\Psi^{0}(1)+\frac{\epsilon^{ABCD}d_{A}d_{B}d_{C}d_{D}}{48\hat{1}^2}\Psi^{1}(1)+\frac{\epsilon^{ABCD}d_{A}d_{B}d_{C}d_{D}}{48\hat{1}^2}\Psi^{2}(1)+\cdots.
\end{split}\end{equation}
Since all the terms $\Psi^{n-2}$ are linearly independent, we have
\begin{equation}\begin{split}
\nonumber
\bar{\Psi}^{0}=\bar{\Phi}&=\frac{\epsilon^{ABCD}d_{A}d_{B}d_{C}d_{D}}{48\hat{1}^2}\Phi\\
 \bar{\Psi}^{1}&=\frac{\epsilon^{ABCD}d_{A}d_{B}d_{C}d_{D}}{48\hat{1}^2}\Psi^{1}\\
\bar{\Psi}^{2}&=\frac{\epsilon^{ABCD}d_{A}d_{B}d_{C}d_{D}}{48\hat{1}^2}\Psi^{2}\\
&\vdots
\end{split}\end{equation}
and so on, thus showing that all the $\Psi^{n-2}$ individually satisfy both the constraints and that they have the same form as (\ref{eq:sfield})
\begin{equation}\begin{split}
\label{eq:multiplessfield}
\Psi(y,\theta)=&\frac{1}{\hat{\partial}}\underline{A}(y)+\frac{i}{\hat{\partial}}\theta^A\underline{\lambda}_A(y)+i\frac{1}{\sqrt{2}}\theta^A\theta^B \underline{\bar{C}}_{AB}(y)\\&+\frac{\sqrt{2}}{3!}\theta^A\theta^B\theta^C\epsilon_{ABCD}\underline{\bar{\lambda}}^D(y)+\frac{1}{12}\theta^A\theta^B\theta^C\theta^D\epsilon_{ABCD}\hat{\partial}\underline{\bar{A}}(y)
\end{split}\end{equation}
where the underscores attached to the component fields are present to distinguish them from the fields present in (\ref{eq:sfield}) and depend on some multiples of the fields in (\ref{eq:sfield}) and we have dropped the superscripts on $\Psi$.\footnote{It is actually a basic fact that products of superfields are also superfields, but as our field $\Psi$ as defined by (\ref{eq:Psi}) consists of fields multiplied together at different points and knitted together with a non-local kernel and integrated over, the situation is not as simple, but as we have discussed it still holds.} For example,
\begin{equation}\begin{split}
\nonumber
\underline{A}_{1}&=\int_{2\cdots n}\frac{-(i)^n\hat{1}^{n-1}}{(1,2)(1,2+3)\cdots (1,2+\cdots+(n-1))} A_{\bar{2}}\cdots A_{\bar{n}}\\
(\underline{\lambda}_{A})_{1}&=\sum_{k=2}^{n}\int_{2\cdots n}\frac{-(i)^n\hat{1}^{n-1}}{(1,2)(1,2+3)\cdots (1,2+\cdots+(n-1))} A_{\bar{2}}\cdots(\lambda_A)_{\bar{k}}\cdots A_{\bar{n}}\\
&\vdots
\end{split}\end{equation}
and so on for $\underline{C},\ \underline{\bar{\lambda}}$ and $\underline{\bar{A}}$. This is a result we need to use later.

\section{Symmetries in Free Supersymmetric Theories}
\label{sec:free}
\subsection{Transformations of N=1 Chiral Free SUSY}
As a precursor to studying symmetries of the N=4 Super Yang-Mills multiplet, let us study a simpler theory with action
\begin{equation}
	S=\int d^4 x\left\{\eta^{\mu\nu}\partial_{\mu}\widetilde{\varphi}\partial_{\nu}\varphi+\widetilde{\psi} i\gamma^{\mu}\partial_{\mu}\psi\right\}.
\end{equation}
Clearly this will be invariant under the component SUSY transformations, for example see \cite{bagger}. If $x\rightarrow x^G$ is a member of the isometry group of the space-time then the action is invariant under
\begin{align}
\nonumber
	\nonumber \delta\varphi(x)&=\epsilon \widetilde{\lambda}\psi(x_G),& \delta\widetilde{\varphi}(x)&=\epsilon\widetilde{\psi}(x_{G^{-1}})\lambda\\
	\delta \psi(x)&=-i\gamma^{\mu}\partial_{\mu}\varphi(x_{G})\lambda,&\delta\widetilde{\psi}(x)&=\widetilde{\lambda} i \gamma^{\mu}\partial_{\mu}\widetilde{\varphi}(x_{G^{-1}})
\end{align}
where $\lambda$ is a constant spinor and the $\gamma^{\mu}$ are the Dirac gamma matrices. This is simple to prove, we have
\begin{equation}\begin{split}
\nonumber
	\nonumber\delta S=\int d^4 x\bigg\{&
	\nonumber\eta^{\mu\nu}\partial_{\mu}\widetilde{\varphi}(x)\partial_{\nu}(\widetilde{\lambda}\psi(x_G))+\frac{1}{2}\eta^{\mu\nu}\partial_{\mu}(\widetilde{\psi}(x_{G^{-1}})\lambda)\partial_{\nu}\varphi(x)\\
	\nonumber&+\widetilde{\psi}(x)i\gamma^{\mu}\partial_{\mu}(-i\gamma^{\nu}\partial_{\nu}\varphi(x_{G})\lambda)+\frac{1}{2}\widetilde{\lambda} i \gamma^{\nu}\partial_{\nu}\widetilde{\varphi}(x_{G^{-1}}))i\gamma^{\mu}\partial_{\mu}\psi(x)\bigg\}.	
\end{split}\end{equation}
Multiplying out terms and then taking $\partial_{\nu}$ out of the third term as a total derivative, we get
\begin{equation}\begin{split}
\nonumber
	\nonumber \delta S=\int d^4 x\bigg\{& \eta^{\mu\nu}\widetilde{\lambda}\partial_{\mu}\widetilde{\varphi}(x)\partial_{\nu}\psi(x_G)-\frac{1}{2}\widetilde{\lambda}\partial_{\mu}\widetilde{\varphi}(x_{G^{-1}})\left\{\gamma^{\mu},\gamma^{\nu}\right\}\partial_{\nu}\psi(x)\\
	\nonumber &+\eta^{\mu\nu}\partial_{\mu}\widetilde{\psi}(x_G^{-1})\partial_{\nu}\varphi(x)\lambda-\frac{1}{2}\partial_{\mu}\widetilde{\psi}(x)\left\{\gamma^{\mu},\gamma^{\nu}\right\}\partial_{\nu}\varphi(x_G)\lambda\bigg\}.
\end{split}\end{equation}
Applying the isometry $x\rightarrow x^G$ in the second and third terms and using the Dirac algebra
\begin{equation}
\nonumber
	\eta^{\mu\nu}=\frac{1}{2}\left\{\gamma^{\mu},\gamma^{\nu}\right\}
\end{equation}
we get $\delta S=0$. 
\subsection{Transformation of the Free N=4 Multiplet}
The free action on the RHS of (\ref{eq:Ssd}) was written as
\begin{equation}
\label{eq:free superfield action}
S=tr\int d^4 x d^4\theta \chi(x,\theta)\Omega(x)\chi(x,\theta)
\end{equation}
where $\Omega(x)=\hat{\partial}\check{\partial}-\tilde{\partial}\bar{\partial}$ and the change in this action is
\begin{equation}
\nonumber
\delta S=2tr\int d^4 x d^4\theta \chi(x,\theta)\Omega(x)\delta\chi(x,\theta)
\end{equation}
where the superfield $\chi(x,\theta)$ is given by (\ref{eq:free_superfield}). The expression for $\delta\chi$ is
\begin{equation}\begin{split}
\nonumber
\delta\chi(y,\theta)=&\frac{1}{\hat{\partial}}\delta B(y)+\frac{i}{\hat{\partial}}\theta^A\delta \rho_A(y)+i\frac{1}{\sqrt{2}}\theta^A\theta^B \delta\bar{D}_{AB}(y)\\&+\frac{\sqrt{2}}{3!}\theta^A\theta^B\theta^C\epsilon_{ABCD}\delta\bar{\rho}^D(y)+\frac{1}{12}\theta^A\theta^B\theta^C\theta^D\epsilon_{ABCD}\hat{\partial}\delta\bar{B}(y)
\end{split}\end{equation}
where $\delta A, \delta\rho, \delta C,\delta\bar{\rho}$ and $\delta\bar{B}$ are to be determined. In component form the free action (\ref{eq:free superfield action}) is easily expanded out to give
\begin{equation}
\label{eq:component free action}
S=tr\int d^4 x \big\{\bar{B}(x)\Omega(x) B(x)+\frac{1}{4}\bar{D}_{AB}(x)\Omega(x) {D}^{AB}(x)+\frac{i}{\sqrt{2}}\bar{\rho}^A(x)\frac{\Omega(x)}{\hat{\partial}}\rho_A(x)\big\}.
\end{equation}

In the paper \cite{brink} they give the supersymmetry transformations of the component fields, $B(x)$, $\rho(x)$ and $D_{AB}(x)$. The transformations of their conjugates can easily be calculated directly from eqns (3.15), (3.16) and (3.17) in their paper, or using the super-symmetry generators $q^A$ and $\bar{q}_A$ given in (\ref{eq:susy gen}). These transformations on their own do indeed leave the free action invariant but we can go further than that. We can write the transformations as
\begin{align}
\label{eq:n=4 susy component transforms}
\nonumber\delta B&=\varepsilon\xi^A\rho_A(x^G)\\
\nonumber\delta\rho_A&=\varepsilon\sqrt{2}\hat{\partial}\bar{D}_{AB}(x^G)\xi^B+\epsilon\sqrt{2}\bar{\xi}_A\hat{\partial}B(x^G)\\
\nonumber\delta D^{AB}&=-i\varepsilon\left(\xi^A\bar{\rho}^B(x^{G^{-1}})-\xi^B\bar{\rho}^A(x^{G^{-1}})+\epsilon^{ABCD}\rho_C(x^G)\bar{\xi}_D\right)\\
\nonumber\delta\bar{D}_{AB}&=-i\varepsilon\left(\rho_A(x^G)\bar{\xi}_B-\rho_B(x^G)\bar{\xi}_A+\epsilon_{ABCD}\xi^C\bar{\rho}^D(x^{G^{-1}})\right)\\
\nonumber\delta\bar{\rho}&=\varepsilon\sqrt{2}\bar{\xi}_B\hat{\partial}D^{BA}(x^{G^{-1}})+\epsilon\sqrt{2}\xi^A\hat{\partial}\bar{B}(x^{G^{-1}})\\
\delta\bar{B}&=-i\varepsilon\bar{\rho}^A(x^{G^{-1}})\bar{\xi}_A,
\end{align}
where $\xi^A$ are finite Grassman numbers carrying $SU(4)$ indices. It is simple to check that the transformations $B\rightarrow B+\delta B, \rho\rightarrow \rho+\delta\rho, \cdots$ with the $\delta$s as given above leave the free action (\ref{eq:component free action}) invariant using the fact $x\rightarrow x^G$ is an isometry, which implies $\Omega(x^G)=\Omega(x)$ and further that the Jacobian of the transformation is unity. 
 However defining the transformations in component form in this manner leads to complications. The terms at the front of the superfield (\ref{eq:free_superfield}), $B$ and $\rho$, are defined to transform  under the isometry $x\rightarrow x_G$ whereas those at the end of the superfield, namely $\bar{\rho}$ and $\bar{A}$, transform under the inverse of the isometry, $x\rightarrow x_{G^{-1}}$. This presents a problem in constructing a superfield formulation of these transformations. It is solved by noticing that we can interchange $x^G$ and $x^{G^{-1}}$ in (\ref{eq:n=4 susy component transforms}) and this will also be a symmetry of the action since we can write $H=G^{-1}$ and do the same calculation. Further, since both these are symmetries, we can add them together to also form a symmetry of the action thus,
\begin{align}
\nonumber\delta B&=\epsilon\xi^A\rho_A(x^G)+\epsilon\xi^A\rho_A(x^{G^{-1}})\\
\nonumber &\vdots\\
\delta\bar{B}&=-i\epsilon\bar{\rho}^A(x^{G^{-1}})\bar{\xi}_A-i\epsilon\bar{\rho}^A(x^{G})\bar{\xi}_A
\end{align}
and then this can be written as the sum of two transformed superfields, one with arguments $x^G$ in the component fields and the other with arguments $x^{G^{-1}}$, so roughly speaking
\begin{equation}
\label{eq:delta chi}
\delta\chi=\varepsilon\chi^G(x)+\varepsilon\chi^{G^{-1}}(x)
\end{equation}
with 
\begin{align}
\nonumber
\label{eq:G transformation}
 \chi^G=&\frac{i\xi^A \rho(x^G)}{\hat{\partial}}+\cdots\\
&-\frac{i}{12}\epsilon^{ABCD}\theta_A\theta_B\theta_C\theta_D\hat{\partial}\bar{\rho}^E(x^G)\bar{\xi}_E
\end{align}
and similarly for $\chi^{G^{-1}}$. The above is simply the susy transformed field with the arguments of the component fields being $x^G$ (or $x^{G^{-1}}$) instead of $x$.

\section{Transformation that leaves N=4 SYM action invariant}
\label{sec:int}
In the paper \cite{me}, the authors calculate symmetries of the non-supersymmetric Chalmers-Siegel action. Given the gauge fields $A$ and $\bar{A}$ they use the field redefinition $A[B]$ mapping the Chalmers-Siegel action to that of the free theory to calculate an expression for $\delta A$ in terms of the free field $\delta B$ order by order in $B$. The inverse expression $B[A]$ is then substituted to arrive at an order by order expansion of $\delta A$ in terms of the $A$ field itself to non-trivial order in perturbation theory. They then guess the expression for $\delta A$ to all orders in perturbation theory and prove that the change in the action is indeed zero. The expression they arrive at for $\delta A$ is
\begin{eqnarray}
\label{eq:deltaA}
	\nonumber\delta A_{1}=-\varepsilon\sum_{n=2}^{\infty}\sum_{i=2}^{n}\sum_{j=i}^n\int_{2\cdots n}\frac{\hat{1}}{\hat{q}}\Gamma(q^G,i^G,\cdots,j^G)\Gamma(q,j+1,\cdots,n,1\cdots,i-1)\times\\
\times A_{2}\cdots A_{\bar{i}^G}\cdots A_{\bar{j}^G}\cdots A_{\bar{n}}
\end{eqnarray} 
where $\Gamma$ is given by 
\begin{equation}
\nonumber
\Gamma(1\cdots n)=-(i)^n\frac{\hat{1}}{(1,2)}\frac{\hat{1}}{(1,2+3)}\cdots\frac{\hat{1}}{(1,2+\cdots (n-1))}.
\end{equation}	

We shall extend this to the supersymmetric Chalmers-Siegel action describing the self-dual sector of $N=4$ supersymmetric Yang-Mills on the light cone (\ref{eq:Ssd}) by following the same procedure described in \cite{me} for $\Delta \Phi$ using (\ref{eq:phi(chi)}) and (\ref{eq:chi(phi)}), and the discussion in \textsection(\ref{sec:free}). We shall guess the expression to all orders in perturbation theory by comparison with (\ref{eq:deltaA}) and substitute this back into the self-dual part of the action (\ref{eq:Ssd}) to prove it does indeed leave the action invariant.

Start with our expression for the field redefinition $\Phi[\chi]$ from (\ref{eq:phi(chi)}). The change in $\Phi$ is then written
\begin{equation}
\nonumber
\Delta\Phi_{1}=\sum_{n=2}^{\infty}\sum_{i=2}^{n}\int_{2\cdots n}C(12\cdots n)\chi_{\bar{2}}\cdots\delta\chi_{\bar{i}}\cdots\chi_{\bar{n}}
\end{equation}
where $\delta\chi$ is the transformation of the free field as defined by (\ref{eq:delta chi}). We use a capital Delta to represent the change in $\Phi$ to distinguish it from the change in the free field $\delta\chi$. Now as discussed, each term in the series expansion is itself a superfield of the form (\ref{eq:sfield}) and 
\begin{equation}
\nonumber
 \delta\chi_{1}=\sum_{n=2}^{\infty}\int_{2\cdots n}\delta\bigg\{D(12\cdots n)\Phi_{\bar{2}}\cdots \Phi_{\bar{n}}\bigg\}.
\end{equation}
We shall expand order by order and collect terms. We have
\begin{eqnarray}\begin{split}
\label{eq:delphi in terms of chi first 4 orders}
\Delta \Phi_{1}=&\varepsilon \delta\chi_{1}+\varepsilon \int_{23}C(123)\left\{\delta\chi_{\bar{2}}\chi_{\bar{3}}+\chi_{\bar{2}}\delta\chi_{\bar{3}}\right\}\\
&+\varepsilon\int_{234}C(1234)\left\{\delta\chi_{\bar{2}}\chi_{\bar{3}}\chi_{\bar{4}}+\chi_{\bar{2}}\delta\chi_{\bar{3}}\chi_{\bar{4}}+\chi_{\bar{2}}\chi_{\bar{3}}\delta\chi_{\bar{4}}\right\}\\
&+\varepsilon\int_{2345}C(12345)\big\{\delta\chi_{\bar{2}}\chi_{\bar{3}}\chi_{\bar{4}}\chi_{\bar{5}}+\chi_{\bar{2}}\delta\chi_{\bar{3}}\chi_{\bar{4}}\chi_{\bar{5}}+\\
&\ \ \ \ \ \ \ \ \ \ \ \ \ \ \ \ \ \ \ \ \ \ \ \ +\chi_{\bar{2}}\chi_{\bar{3}}\delta\chi_{\bar{4}}\chi_{\bar{5}}+\chi_{\bar{2}}\chi_{\bar{3}}\chi_{\bar{4}}\delta\chi_{\bar{5}}\big\}\\
&+\cdots.
\end{split}\end{eqnarray}
Now, as per \cite{me} we substitute the inverse expression (\ref{eq:chi(phi)}) into the above to get the extremely cumbersome expression given by (\ref{eq:delta Phi expansion}). Recall that we wrote down the inverse of (\ref{eq:phi(chi)}) as (\ref{eq:chi(phi)}). Collecting like terms and writing their coefficients in terms of their independent momenta the expression reduces nicely. We shall write it out order by order here, where the argument in the kernels labelled with a $(-)$ is taken to be minus the sum of the remaining arguments. First order is simply $\Delta\Phi=\delta\Phi+\cdots$.
\paragraph{Second Order}
\begin{equation}
\label{eq:O(2)}
\cdots-\varepsilon\int_{23}\bigg\{\frac{\hat{q}}{\hat{1}}\delta\left\{D(-\ 23)\Phi_{\bar{2}}\Phi_{\bar{3}}\right\}+\frac{\hat{q}}{\hat{1}}\delta\Phi_{\bar{2}}D(-\ 31)\Phi_{\bar{3}}+\frac{\hat{q}}{\hat{1}}D(-\ 12)\Phi_{\bar{2}}\delta\Phi_{\bar{3}}\bigg\}
\end{equation}
\paragraph{Third Order}
\begin{equation}\begin{split}
\label{eq:O(3)}
 \cdots-\varepsilon\int_{234}\bigg\{&\frac{\hat{q}}{\hat{1}}\delta\left\{D(-\ 234)\Phi_{\bar{2}}\Phi_{\bar{3}}\Phi_{\bar{4}}\right\}+\frac{\hat{q}}{\hat{1}}\delta\left\{D(-\ 23)\Phi_{\bar{2}}\Phi_{\bar{3}}\right\}D(-\ 41)\Phi_{\bar{4}}\\
&+\frac{\hat{q}}{\hat{1}}D(-\ 12)\Phi_{\bar{2}}\delta\left\{D(-\ 34)\Phi_{\bar{3}}\Phi_{\bar{4}}\right\}+\frac{\hat{q}}{\hat{1}}\delta\Phi_{\bar{2}}D(-\ 341)\Phi_{\bar{3}}\Phi_{\bar{4}}\\
&\frac{\hat{q}}{\hat{1}}\Phi_{\bar{2}}\delta\Phi_{\bar{3}}D(-\ 412)\Phi_{\bar{4}}+\frac{\hat{q}}{\hat{1}}D(-\ 123)\Phi_{\bar{2}}\Phi_{\bar{3}}\delta\Phi_{\bar{4}}\bigg\}
\end{split}\end{equation}
The fourth order expression is written down in appendix (A).
Note that in the above, for each term containing $\delta\left\{D(q,i,\cdots ,j)\Phi(i)\cdots \Phi(j)\right\}$, we define $q$ to be $q=p_i+\cdots +p_j$. As was done in \cite{me} we now hypothesize a generalisation to the expression given in that paper, (eqn (4.4) in their paper). We write
\begin{equation}\begin{split}
\label{eq:DeltaPhi}
 \Delta\Phi_{1}=-\varepsilon\sum_{n=2}^{\infty}\sum_{i=2}^{n}\sum_{j=i}^{n}\int_{2\cdots n}\frac{\hat{q}}{\hat{1}}\Phi_{\bar{2}}\cdots\Phi_{\overline{i-1}}\delta\left\{D(-,i,\cdots,j)\Phi_{\bar{i}}\cdots\Phi_{\bar{j}}\right\}\times&\\
\times D(-,j+1,\cdots,n,1,\cdots,i-1)\Phi_{\overline{j+1}}\cdots\Phi_{\bar{n}}.&
\end{split}\end{equation}
It is now a simply a matter of proving that this expression leaves the Chalmers-Siegel action of self-dual Yang-Mills (\ref{eq:Ssd}) invariant which is a similar calculation to that in \cite{me}. Because of the CPT self conjugacy property of the N=4 SYM multiplet, then the ensuing calculation is in fact easier than that given in the pure Yang-Mills setting of \cite{me} as $\Delta\bar{\Phi}$ is eliminated from the self-dual part of the action.
\begin{figure}[h]
\begin{center}
    \includegraphics{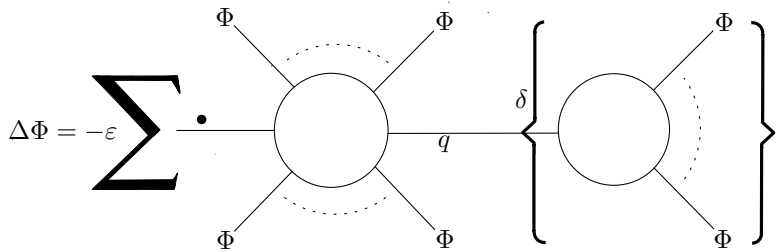}
    \caption{$\Delta \Phi$}
 \end{center}
\end{figure}
Figure (1) is a diagrammatic expression of (\ref{eq:DeltaPhi}) with the the dotted leg representing the argument not integrated over, i.e. $p_1$. In the paper \cite{me}, the transformation of the free fields was written as $\delta B(p)=\varepsilon B(p_G)$ and in the expression for $\delta A$ they wrote something of the form $A_{\bar{2}}\cdots A_{\overline{i-1}}A_{\bar{i}^G}\cdots A_{\bar{j}^G}A_{\overline{j+1}}\cdots A_{\bar{n}}$. Here the situation is more complicated and an operation is performed on the group of fields enclosed in the parenthesise that mixes up fermionic and bosonic degrees of freedom. The summation here means to sum over all $n$, the total number of legs, and all $i$ and $j$ with $2\leq i \leq j \leq n$. As we shall see later, there is no conceptual difficulty in calculating the transformations of each of the component fields.

First however, let us now consider how to prove that the transformation $\Phi\rightarrow\grave{\Phi}=\Phi+\Delta\Phi$ does indeed leave the action (\ref{eq:Ssd}) invariant. The change in the action is
\begin{equation}\begin{split}
\nonumber
 \Delta S_{SD}=&2tr\int d^4 p d^4 \theta \ \ \Phi(p)\Omega(p)\Delta\Phi(-p)+\\
&+\frac{2}{3}tr\int_{123}d^4 \theta \ \ \ \hat{p}_1 \left\{\bar{p}_3-\bar{p}_2\right\}\Delta\Phi(-1)\Phi(-2)\Phi(-3)\\
&+\frac{2}{3}tr\int_{123}d^4 \theta \ \ \ \hat{p}_1 \left\{\bar{p}_3-\bar{p}_2\right\}\Phi(-1)\Delta\Phi(-2)\Phi(-3)\\
&+\frac{2}{3}tr\int_{123}d^4 \theta \ \ \hat{p}_1\left\{\bar{p}_3-\bar{p}_2\right\}\Phi(-1)\Phi(-2)\Delta\Phi(-3)
\end{split}\end{equation}
with $\Omega(p)=\hat{p}\check{p}-\tilde{p}\bar{p}$ as before, after transforming into momentum space and stripping off $\delta$ functions and various factors of $2\pi$. Using momentum conservation and the cyclical property of the trace, (recall that the fields contain the generators of the gauge group), and relabelling arguments the change in the action easily reduces to
\begin{equation}\begin{split}
\nonumber
 \Delta S_{SD}=&2tr\int d^4 p d^4 \theta \ \ \Phi(p)\Omega(p)\Delta\Phi(-p)+\\
&-2 tr\int_{12}d^4 \theta \ \ \ \ \left\{p_1,p_2\right\}\Delta\Phi(1+2)\Phi(-1)\Phi(-2)
\end{split}\end{equation}
where the bracket $\left\{\ ,\ \right\}$ is defined as $\left\{p_i,p_j\right\}=\hat{p}_i \bar{p}_j-\bar{p}_i \hat{p}_j$. We shall separate the calculation into two distinct parts, $\Delta S_F$, the free part and $\Delta S_I$, the interaction. The diagrams for these are given in fig (2), clearly the free part is just a two point vertex with $\Omega$ as the vertex factor which as we recall is invariant under the isometry $x\rightarrow x^G$ and the interacting part is a 3 point vertex with I given by $\left\{k,k+1\right\}$. The vertex factor $I$ is clearly not invariant under $x\rightarrow x^G$ For each part we shall collect all possible diagrams and extract algebraic expressions from them for $\Delta S_F$, the free part, and $\Delta S_I$ and show that $\Delta S_{SD}=\Delta S_F-\Delta S_I=0$.
\begin{figure}[h]
\begin{center}
    \includegraphics{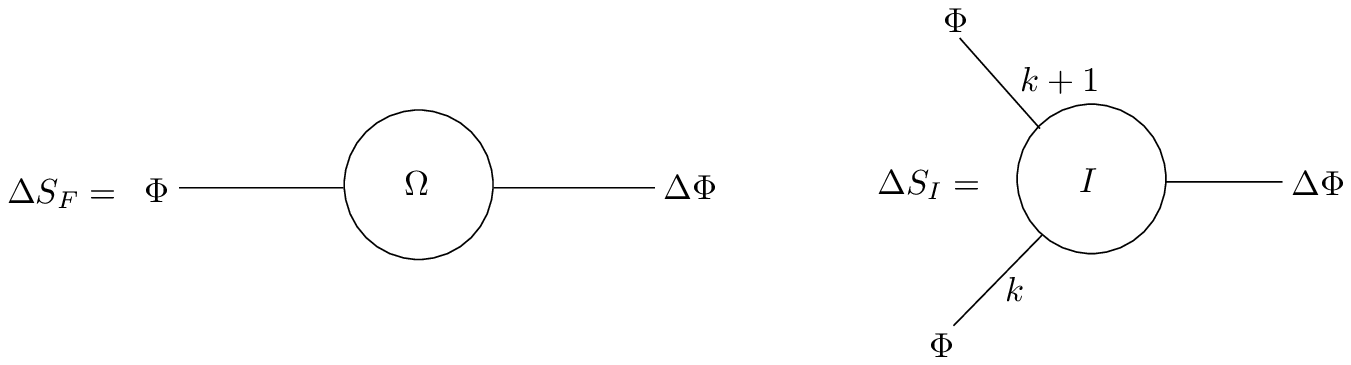}
    \caption{Change in the self-dual action $\Delta S$}
 \end{center}
\end{figure}
As discussed earlier, the expression enclosed in the brackets
\begin{equation}
\nonumber
 \Psi_q=\int_{i\cdots j}D(q,i,\cdots ,j)\Phi_{\bar{i}}\cdots \Phi_{\bar{j}}
\end{equation}
is itself a superfield with argument $q$ satisfying (\ref{eq:constraint1}) and (\ref{eq:constraint2}) and the solutions to the constraints are expressed in (\ref{eq:multiplessfield}). The transformation (\ref{eq:delta chi}) is applied to the superfield enclosed in the brackets with component fields $\underline{A}$, $\underline{\lambda}$, $\underline{C}$, $\underline{\bar{\lambda}}$ and $\underline{\bar{A}}$.  This is represented as a diagram in fig (3).
\begin{figure}[h]
\begin{center}
    \includegraphics{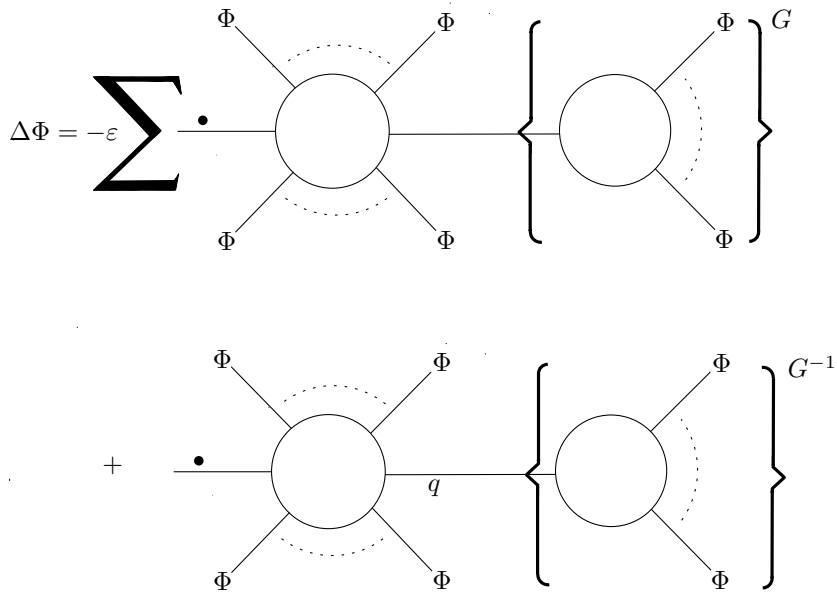}
    \caption{$\Delta \Phi$}
 \end{center}
\end{figure}
We will proceed by drawing all the possible diagrams that make up $\Delta S_F$, fig (4). Now the expression in brackets satisfies the constraints, and so does the part outside the brackets. We write 
\begin{equation}\begin{split}
\nonumber
 \Delta S_F=-\varepsilon\sum_{n=2}^{\infty}\sum_{i=2}^{n}\sum_{j=i}^{n}tr\int_{12\cdots n}\frac{\hat{q}}{\hat{1}}\Omega(1)\Phi_{\bar{1}}\Phi_{\bar{2}}\cdots\Phi_{\overline{i-1}}\left\{D(-,i,\cdots,j)\Phi_{\bar{i}}\cdots\Phi_{\bar{j}}\right\}^G\times&\\
\times D(-,j+1,\cdots,n,1,\cdots,i-1)\Phi_{\overline{j+1}}\cdots\Phi_{\bar{n}}&\\
+G\rightarrow G^{-1}.
\end{split}\end{equation}
Next we can use the cyclicity of the trace and relabel arguments as follows 
\begin{equation}\begin{split}
\nonumber
 \Delta S_F=-\varepsilon\sum_{n=2}^{\infty}\sum_{j=1}^{n-1}\sum_{k=j+1}^{n}tr\int_{12\cdots n}\frac{\hat{q}}{\hat{k}}\Omega(k)\left\{D(-,1,\cdots,j)\Phi_{\bar{1}}\cdots\Phi_{\bar{j}}\right\}^G\times&\\
\times D(-,j+1,\cdots,n)\Phi_{\overline{j+1}}\cdots\Phi_{\bar{n}}&\\
+G\rightarrow G^{-1}
\end{split}\end{equation}
and then define 
\begin{align}
\nonumber
 \Psi&=\int_{1\cdots j}D(q,1,\cdots,j)\Phi_{\bar{1}}\cdots\Phi_{\bar{j}}\\
\nonumber
\Theta&=\int_{j+1\cdots n}\frac{\hat{q}\Omega(k)}{\hat{k}}D(q,j+1,\cdots,n)\Phi_{\overline{j+1}}\cdots\Phi_{\bar{n}}.
\end{align}
Both of these have the form (\ref{eq:Psi}) and as we proved in \textsection(\ref{sec:mhvruled}) satisfy the constraints (\ref{eq:constraint1}) and (\ref{eq:constraint2}) and the solution to these constraints are of the form (\ref{eq:multiplessfield}) with a similar expression for $\Theta$. By writing in component form it is possible to show that
\begin{equation}
\nonumber
 \Delta S_F=tr\int_{q}\Psi^{G^{-1}}(q)\Theta(q)=-tr\int_{q}\Psi(q)\Theta^G(q).
\end{equation}
by utilizing the integral over $\theta$ which picks out the $\theta^4$ component. 
\begin{figure}[h]
\begin{center}
    \includegraphics{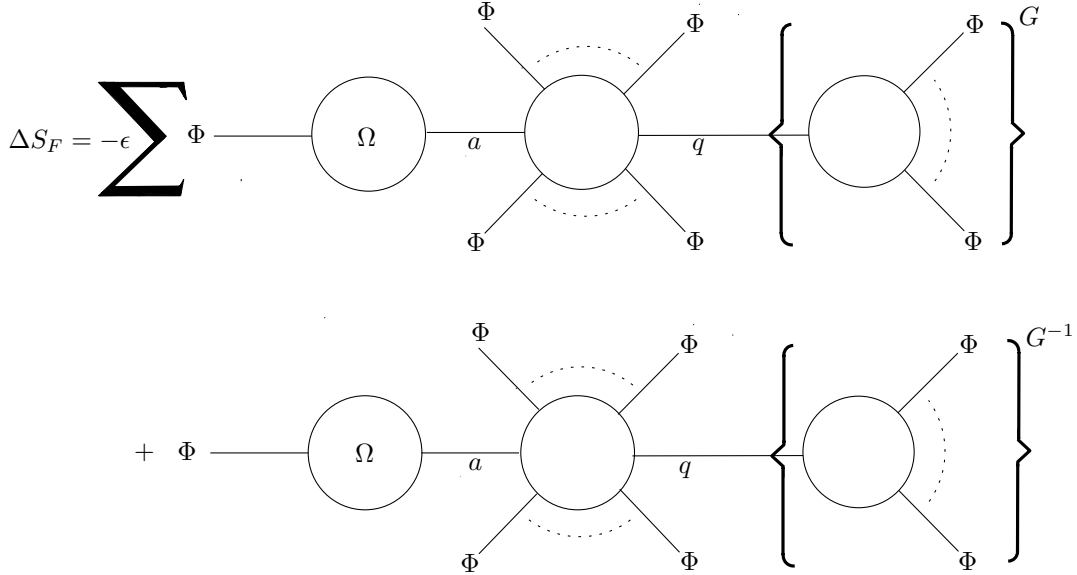}
    \caption{Change in the free part of the self-dual action $\Delta S_F$}
 \end{center}
\end{figure}
Therefore, fig (4) becomes fig (5) and there is a summation over cyclic rotations of the vertex $\Omega$ around the diagram. The diagram of fig (5) becomes the following expression where a factor of $\left\{D\Phi\cdots\Phi\right\}^G D\Phi\cdots\Phi$ can be taken out of a cyclic clockwise sum over rotations of $\Omega$ with the momentum of the first leg enclosed in brackets being $p_1$,
\begin{equation}
\nonumber
 \Delta S_F=-\varepsilon \sum_{n=2}^{\infty}\sum_{j=1}^{n-1}tr\int_{1\cdots n} X(1,j)\left\{D(-,1,\cdots,j)\Phi_{\bar{1}}\cdots\Phi_{\bar{j}}\right\}^GD(-,j+1,\cdots,n)\Phi_{\overline{j+1}}\cdots\Phi_{\bar{n}}
\end{equation}
where the coefficient $X_{1,j}$ is given by
\begin{equation}
\nonumber
X(1,j)=\frac{\hat{q}^G}{\hat{1}^G}\Omega_1^G+\cdots+\frac{\hat{q}^G}{\hat{j}^G}\Omega_j^G+\frac{\hat{q}}{\widehat{j+1}}\Omega_{j+1}+\cdots+\frac{\hat{q}}{\hat{n}}\Omega_n.
\end{equation}
\begin{figure}[h]
\begin{center}
    \includegraphics{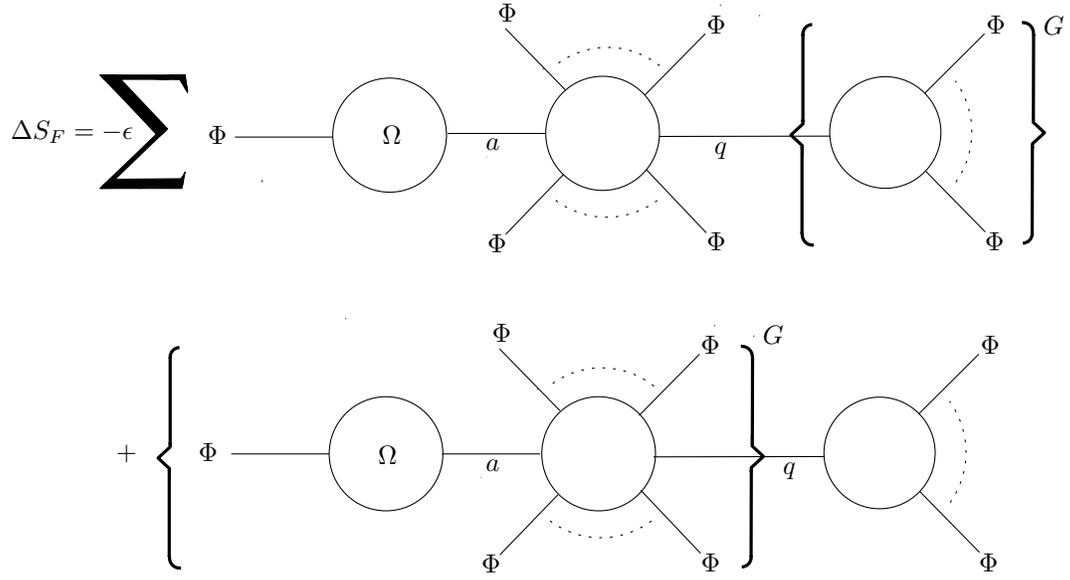}
    \caption{Change in the free part of the self-dual action $\Delta S_F$}
 \end{center}
\end{figure}

Moving on to the cubic part of the action, $\Delta S_I$, we have a sum over cyclic rotations of a three point vertex with factor $I$ around all possible diagrams, as given in fig (6). We can then undo the transformation in the second diagram as before to arrive at fig (7). Similarly then, rotating the vertex $I$ clockwise around the diagram and with the first leg enclosed in the brackets has momentum $p_1$, we have
\begin{figure}[h]
\begin{center}
    \includegraphics{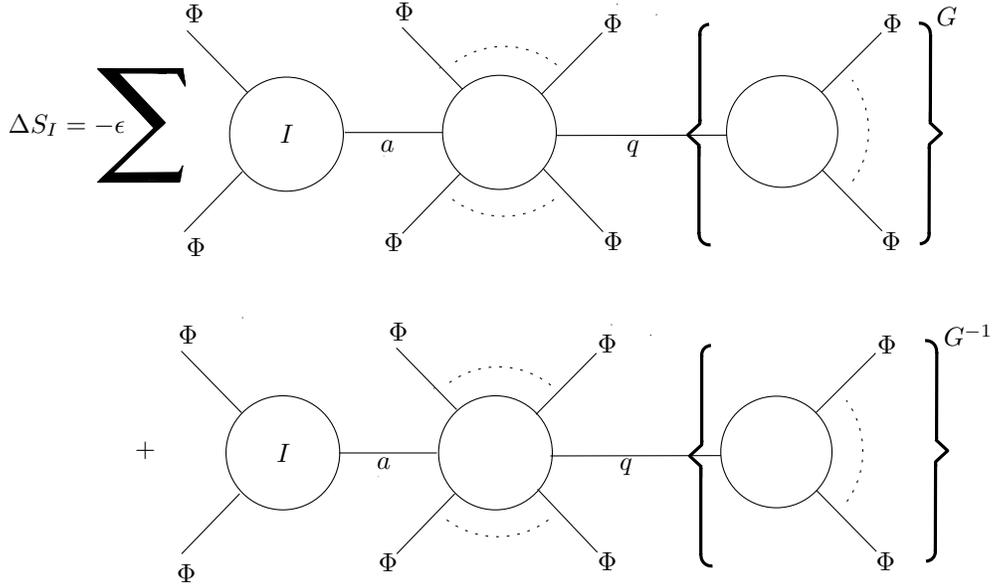}
    \caption{Change in the interacting part of the self-dual action $\Delta S_I$}
 \end{center}
\end{figure}

\begin{equation}
\nonumber
 \Delta S_I=-\varepsilon\sum_{n=2}^{\infty}\sum_{j=1}^{n-1}tr\int_{1\cdots n} Y(1,j)\left\{D(-,1,\cdots,j)\Phi_{\bar{1}}\cdots\Phi_{\bar{j}}\right\}^G D(-,j+1,\cdots,n)\Phi_{\overline{j+1}}\cdots\Phi_{\bar{n}}
\end{equation}
with
\begin{equation}
\label{eq:SI 10}
Y(1,j)=-\sum_{k=1}^{j-1}\frac{\{\hat{k}^G,(k+1)^G\}}{k^G,\widehat{k+1}^G}(q,p_{1,k})-\sum_{k=j+1}^{n-1}\frac{\{k,k+1\}}{\hat{k}\widehat{k+1}}(q,p_{j+1,k}).
\end{equation}
\begin{figure}[h]
\begin{center}
    \includegraphics{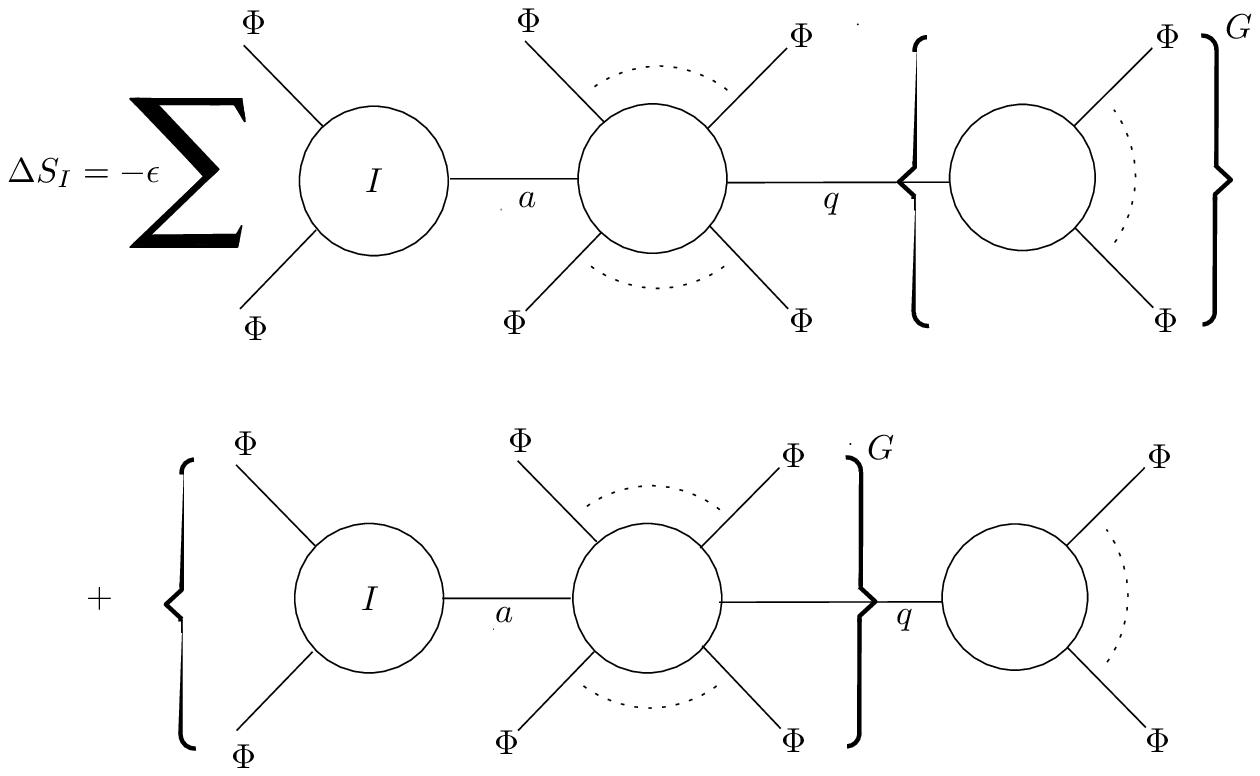}
    \caption{Change in the interacting part of the self-dual action $\Delta S_I$}
 \end{center}
\end{figure} 
The left hand sum should be interpreted as zero when $j=1$ and the right hand sum should be interpreted as zero when $j=n-2$. When expanded out, the summations reduce to 
\begin{equation}	\nonumber\sum_{k=i}^{j-1}\frac{\left\{k,k+1\right\}}{\hat{k}\widehat{k+1}}\left(q,P_{i,k}\right)=-\hat{q}\left(\omega_{-q}+\omega_i+\cdots+\omega_j\right)
\end{equation}
where $\omega_p=\bar{p}\tilde{p}/\hat{p}$. Evaluating the sums in (\ref{eq:SI 10}) we get
\begin{equation}\nonumber Y_{1,j}=\Big(\hat{q}^G\Big\{\omega_{-q}^G+\omega_1^G+\cdots+\omega_j^G\Big\}+\hat{q}\Big\{\omega_{q}+\omega_{j+1}+\cdots+\omega_{n}\Big\}\Big)
\end{equation} 
Now, $-q+p_1+\cdots+p_j=0$ and $-p_{j+1}-\cdots -p_n=0$ we subtract these from each of the above brackets to arrive at
\begin{equation}\begin{split}	 \nonumber Y_{1,j}=\hat{q}^G\Big\{\omega_{-q}^G-\widecheck{-q}^G+\omega_1^G-\check{1}^G&+\cdots+\omega_j^G-\check{j}^G\Big\}+\\	&+\hat{q}\Big\{\omega_{q}-\check{q}+\omega_{j+1}-\widecheck{j+1}+\cdots+\omega_{n}-\widecheck{n}\Big\}
\end{split}\end{equation}
and then take out a factor of $1/\hat{p}$ from each term $\omega_p-\check{p}$ as follows
\begin{eqnarray}
	\nonumber	Y_{1,j}=\frac{\hat{q}^G}{\hat{q}^G}\Omega_{q}^G+\frac{\hat{q}^G}{\hat{1}^G}\Omega_1^G+\cdots+\frac{\hat{q}^G}{\hat{j}^G}\Omega_j^G-\frac{\hat{q}}{\hat{q}}\Omega_{-q}+\frac{\hat{q}}{\widehat{j+1}}\Omega_{j+1}+\cdots+\frac{\hat{q}}{\widehat{n}}\Omega_{n}
\end{eqnarray}
Terms in $\Omega_q$ and and $\Omega_{-q}$ cancel since $\Omega^G=\Omega$. We arrive at
\begin{equation}
	\nonumber	Y_{1,j}=\frac{\hat{q}^G}{\hat{1}^G}\Omega_1^G+\cdots+\frac{\hat{q}^G}{\hat{j}^G}\Omega_j^G+\frac{\hat{q}^G}{\widehat{j+1}}\Omega_{j+1}+\cdots+\frac{\hat{q}}{\widehat{n}}\Omega_{n}\\
=X_{i,j}
\end{equation}
and since all terms in the summation over $j$ and $n$ are linearly independent and sum to zero, we arrive at the result, $\Delta S=\Delta S_F-\Delta S_I=0$ as required.

Since we now have calculated an expression for $\Delta \Phi$ and proved it, we can in principle calculate the expressions for the transformations of the component fields. For example, let us pick out the zeroth order $\theta$ component of $\Delta S$. We will have diagrams of the form fig (8)
\begin{figure}[h]
\begin{center}
    \includegraphics{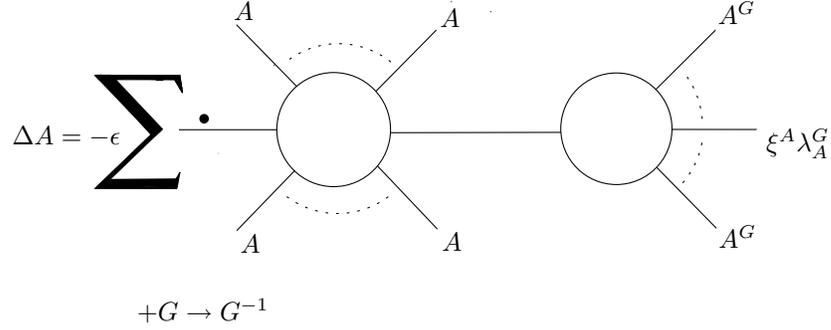}
    \caption{Change in the component field $\Delta A$}
 \end{center}
\end{figure}
or algebraically 
\begin{equation}\begin{split}
 \label{eq:deltaAbar}
\nonumber \delta A_1=-\epsilon\sum_{n=2}^{\infty}\sum_{k=2}^n\sum_{i=2}^{k}\sum_{j=k}^{n}\int_{2 \cdots n}\frac{\hat{1}}{\hat{q}}\Gamma(q,i,\cdots,j)\Gamma(q,j+1,\cdots,n,1,\cdots,i-1)\times&\\
 \times A_{\bar{2}}\cdots A_{\bar{i}^G}\cdots \xi^A\lambda_{A\,\bar{k}^G}\cdots A_{\bar{j}^G}\cdots A_{\bar{n}}&\\
+G\rightarrow G^{-1}.&
\end{split}\end{equation}
Similarly, one could calculate more complicated expressions for the rest of the component fields.

\section{Summary and Conclusions}
We began by reviewing the formulation of light cone N=4 super Yang-Mills theory on the light cone \cite{huang} and writing the action in terms of superfields $\Phi$ and $\bar{\Phi}$, (\ref{eq:N=4 superfield SYM action}). The CPT self conjugacy property of the fields was used to express the action in terms of $\Phi$ only, at the expense of introducing covariant derivatives in the action. The self-dual part however contains only 4 covariant derivatives giving us the Chalmers-Siegel action \cite{Chalmers:1996rq} which is free classically. The full action contains the wrong helicity content for consistency with the MHV rules so we define a canonical transformation from the self-dual sector to a free theory with field variables $\chi$ and we write down the result given in \cite{huang} for the expression $\Phi[\chi]$, (\ref{eq:phi(chi)}). Further, we calculate the inverse of this field redefinition by writing down an ansatz for $\chi[\Phi]$ and substituting it into a recursion relation to prove it.

It was briefly discussed how to contruct symmetries in $N=1$ chiral supersymmetry using isometries $x\rightarrow x^G$. This helped us to see how to construct symmetries of a free $N=4$ susy theory with action (\ref{eq:free superfield action}). We proceeded to calculate a field transformation order by order (up to fourth order in the field variables) by writing $\Delta\Phi$ in terms of the free fields $\chi$, whose transformation we knew, and then substituted the inverse field redefinition to write $\Delta\Phi$ in terms of the original variables. We used these results to guess an expression to all orders in perturbation theory and this was proved by substituting it back into the self-dual action (\ref{eq:Ssd}) to show $\Delta S_{SD}=0$ thus proving our final result (\ref{eq:DeltaPhi}). We concluded by showing how, in principle, we can use our result to calculate the component field transformations, and gave an example of the simplest calculations by writing down $\Delta A$ in terms of $A$ and $\lambda$.

Symmetries of the self-dual Yang-Mills equations were discussed in the paper \cite{popov-1999-11} from the twistor viewpoint. It may be interesting to extend our methods here to the twistor formalism using the $N=4$ Yang-Mills twistor action introduced in the recent paper \cite{popov-2009}.

\allowdisplaybreaks
\acknowledgments
It is a pleasure to thank the STFC for a studentship. I also wish to thank my PhD supervisor, Paul Mansfield, for valuable support and advice, and Alexander Popov for bringing to my attention the papers \cite{popov-1996-374,popov-1999-11,popov-2009}.
\appendix

\section{Order by Order Calculation of $\Delta \Phi$ up to Quartic Terms}
In \textsection(\ref{sec:int}) we wrote down $\Delta \Phi$ in terms of the free field $\chi$, (\ref{eq:delphi in terms of chi first 4 orders}). Carefully substituting the inverse field redefinition, (\ref{eq:chi(phi)}) being careful to label the arguments correctly and maintain the order of the fields, we arrive at

\begin{align}
\label{eq:delta Phi expansion}
 \nonumber\delta \Phi_{1}=\varepsilon\delta\Phi_{1}&+\varepsilon\int_{23}\delta\left\{D(123)\Phi_{\bar{2}}\Phi_{\bar{3}}\right\}+\int_{234}\delta\left\{D(1234)\Phi_{\bar{2}}\Phi_{\bar{3}}\Phi_{\bar{4}}\right\}\\
\nonumber&\ \ \ \ \ \ \ \ \ \ \ \ \ \ \ \ \ \ \ \ \ \ \ \ \ \ \ \ \ \ \ \ \ \ \ \ \ \ \ \ \ \ \ \ +\varepsilon\int_{23}\delta\left\{D(12345)\Phi_{\bar{2}}\Phi_{\bar{3}}\Phi_{\bar{4}}\Phi_{\bar{5}}\right\}\\
\nonumber\\
\nonumber&+\varepsilon\int_{23}C(123)\bigg(\delta\Phi_{\bar{2}}+\int_{45}\delta\left\{D(-245)\Phi_{\bar{4}}\Phi_{\bar{5}}\right\}+\int_{456}\delta\left\{D(-2456)\Phi_{\bar{4}}\Phi_{\bar{5}}\Phi_{\bar{6}}\right\}\bigg)\times\\
\nonumber&\ \ \ \ \ \ \ \ \ \ \ \ \ \ \ \ \times \bigg(\Phi_{\bar{3}}+\int_{78}D(-378)\Phi_{\bar{7}}\Phi_{\bar{8}}+\int_{789}D(-3789)\Phi_{\bar{7}}\Phi_{\bar{8}}\Phi_{\bar{9}}\bigg)\\
\nonumber\\
\nonumber&+\varepsilon\int_{23}C(123)\bigg(\Phi_{\bar{2}}+\int_{45}D(-245)\Phi_{\bar{4}}\Phi_{\bar{5}}+\int_{456}D(-2456)\Phi_{\bar{4}}\Phi_{\bar{5}}\Phi_{\bar{6}}\bigg)\times\\
\nonumber&\ \ \ \ \ \ \ \ \ \ \ \ \ \ \ \ \times \bigg(\delta\Phi_{\bar{3}}+\int_{78}\delta\left\{D(-378)\Phi_{\bar{7}}\Phi_{\bar{8}}\right\}+\int_{789}\delta\left\{D(-3789)\Phi_{\bar{7}}\Phi_{\bar{8}}\Phi_{\bar{9}}\right\}\bigg)\\
\nonumber\\
\nonumber&+\varepsilon\int_{234}C(1234)\bigg(\delta\Phi_{\bar{2}}+\int_{56}\delta\left\{D(-256)\Phi_{\bar{5}}\Phi_{\bar{6}}\right\}\bigg)\times\\
\nonumber&\ \ \ \ \ \ \ \ \ \ \ \ \ \ \ \ \ \ \times\bigg(\Phi_{\bar{3}}+\int_{78}D(-378)\Phi_{\bar{7}}\Phi_{\bar{8}}\bigg)\bigg(\Phi_{\bar{4}}+\int_{9\ 10}D(-49\ 10)\Phi_{\bar{9}}\Phi_{\bar{10}}\bigg)\\
\nonumber\\
\nonumber&+\varepsilon\int_{234}C(1234)\bigg(\Phi_{\bar{2}}+\int_{56}D(-256)\Phi_{\bar{5}}\Phi_{\bar{6}}\bigg)\times\\
\nonumber&\ \ \ \ \ \ \ \ \ \ \ \ \ \ \ \ \ \ \times\bigg(\delta\Phi_{\bar{3}}+\int_{78}\delta\left\{D(-378)\Phi_{\bar{7}}\Phi_{\bar{8}}\right\}\bigg)\bigg(\Phi_{\bar{4}}+\int_{9\ 10}D(-49\ \nonumber10)\Phi_{\bar{9}}\Phi_{\bar{10}}\bigg)\\
\nonumber\\
\nonumber&+\varepsilon\int_{234}C(1234)\bigg(\Phi_{\bar{2}}+\int_{56}D(-256)\Phi_{\bar{5}}\Phi_{\bar{6}}\bigg)\times\\
\nonumber&\ \ \ \ \ \ \ \ \ \ \ \ \ \ \ \ \ \ \times\bigg(\Phi_{\bar{3}}+\int_{78}D(-378)\Phi_{\bar{7}}\Phi_{\bar{8}}\bigg)\bigg(\delta\Phi_{\bar{4}}+\int_{9\ 10}\delta\left\{D(-49\ 10)\Phi_{\bar{9}}\Phi_{\bar{10}}\right\}\bigg)\\
\nonumber\\
\nonumber&+\int_{2345}C(12345)\delta\Phi_{\bar{2}}\Phi_{\bar{3}}\Phi_{\bar{4}}\Phi_{\bar{5}}+\int_{2345}C(12345)\Phi_{\bar{2}}\delta\Phi_{\bar{3}}\Phi_{\bar{4}}\Phi_{\bar{5}}\\
\nonumber\\
&+\int_{2345}C(12345)\Phi_{\bar{2}}\Phi_{\bar{3}}\delta\Phi_{\bar{4}}\Phi_{\bar{5}}+\int_{2345}C(12345)\Phi_{\bar{2}}\Phi_{\bar{3}}\Phi_{\bar{4}}\delta\Phi_{\bar{5}}.
\end{align}
This is a somewhat cumbersome expression but we proceed by collecting like terms. First order is simply $\Delta\Phi_{1}=\varepsilon\delta\Phi_{1}+\cdots$. Second order gives us
\begin{equation}
\nonumber
 \cdots+\varepsilon\int_{23}\bigg(\delta\left\{D(123)\Phi_{\bar{2}}\Phi_{\bar{3}}\right\}+\delta\Phi_{\bar{2}}C(123)\Phi_{\bar{3}}+C(123)\Phi_{\bar{2}}\delta\Phi_{\bar{3}}\bigg)+\cdots
\end{equation}
and further, when the coefficients $C$ and $D$  are written explicitly in terms of their arguments, rearranged and momentum conservation used this becomes (\ref{eq:O(2)}),
\begin{equation}
\nonumber
\cdots-\varepsilon\int_{23}\bigg\{\frac{\hat{q}}{\hat{1}}\delta\left\{D(-\ 23)\Phi_{\bar{2}}\Phi_{\bar{3}}\right\}+\frac{\hat{q}}{\hat{1}}\delta\Phi_{\bar{2}}D(-\ 31)\Phi_{\bar{3}}+\frac{\hat{q}}{\hat{1}}D(-\ 12)\Phi_{\bar{2}}\delta\Phi_{\bar{3}}\bigg\}.
\end{equation}
We shall now extract the third order terms, carefully multiplying out the brackets and keeping terms cubic in $\Phi$ and then relabelling variables of integration. We arrive at
\begin{equation}\begin{split}
\nonumber
 \cdots+\varepsilon\int_{234}\bigg\{&\delta\left\{D(1234)\Phi_{\bar{2}}\Phi_{\bar{3}}\Phi_{\bar{4}}\right\}+\\
&+C(154)\delta\left\{D(-523)\Phi_{\bar{2}}\Phi_{\bar{3}}\right\}\Phi_{\bar{4}}+C(125)\Phi_{\bar{2}}\delta\left\{D(-534)\Phi_{\bar{3}}\Phi_{\bar{4}}\right\}+\\
&+\delta\Phi_{\bar{2}}\left\{C(125)D(-534)+C(1234)\right\}\Phi_{\bar{3}}\Phi_{\bar{4}}+\\
&+\left\{C(154)D(-523)+C(1234)\right\} \Phi_{\bar{2}}\Phi_{\bar{3}}\delta\Phi_{\bar{4}}+\\
&+C(1234)\Phi_{\bar{2}}\delta\Phi_{\bar{3}}\Phi_{\bar{4}}\bigg\}+\cdots
\end{split}\end{equation}
and the first argument of the kernels $C$ and $D$ is equal to minus the sum of the remaining arguments. For example, in the second term, $-p_5=-p_2-p_3=p_1+p_4$. Now write individual terms in terms of the independent momenta using the expressions (\ref{eq:C}) and (\ref{eq:D}) to arrive at our third order expression 
\begin{equation}\begin{split}
\nonumber
 \cdots+\varepsilon\int_{234}\bigg\{\delta\big\{\frac{-\hat{5}\hat{2}\hat{3}\hat{4}}{(5,2)(5,2+3)}\Phi_{\bar{2}}\Phi_{\bar{3}}\Phi_{\bar{4}}\big\}-\frac{\hat{5}\hat{4}}{(5,4)}\delta\big\{\frac{\hat{2}\hat{3}}{(-5,2)}\Phi_{\bar{2}}\Phi_{\bar{3}}\big\}\Phi_{\bar{4}}\\
-\frac{\hat{2}\hat{5}}{(2,5)}\Phi_{\bar{2}}\delta\big\{\frac{\hat{3}\hat{4}}{(-5,3)}\Phi_{\bar{3}}\Phi_{\bar{4}}\big\}+\delta\Phi_{\bar{2}}\frac{\hat{2}^2\hat{3}\hat{4}}{(2,3)(2,3+4)}\Phi_{\bar{3}}\Phi_{\bar{4}}\\
+\Phi_{\bar{2}}\delta\Phi_{\bar{3}}\frac{\hat{2}\hat{3}^2\hat{4}}{(3,4)(3,4+1)}\Phi_{\bar{4}}+\Phi_{\bar{2}}\Phi_{\bar{3}}\delta\Phi_{\bar{4}}\frac{\hat{2}\hat{3}\hat{4}^2}{(4,1)(4,1+2)}\bigg\}+\cdots.
\end{split}\end{equation}

These calculations are almost identical to those performed in \cite{me} so the reader may wish to check these calculations by referring to this paper to verify that the final expression is indeed (\ref{eq:O(3)}). The fourth order expression is also calculable without a great deal of effort. By writing the fourth order terms out, and then expressing them in terms of independent momenta in the above manner, we find that the calculations are again similar to those in \cite{me} (See the equation immediately before A.1 in that paper) and so proceed with the calculation in the same manner they do to find simpler expressions, and then write in terms of the kernel $D$. We get

\begin{equation}\begin{split}
\nonumber
 \cdots&-\varepsilon\int_{2345}\bigg\{\frac{\hat{q}}{\hat{1}}\delta\left\{D(-\ 2345)\Phi_{\bar{2}}\Phi_{\bar{3}}\Phi_{\bar{4}}\Phi_{\bar{5}}\right\}+\frac{\hat{q}}{\hat{1}}\delta\left\{D(-\ 234)\Phi_{\bar{2}}\Phi_{\bar{3}}\Phi_{\bar{4}}\right\}D(-\ 51)\Phi_{\bar{5}}\\
&+\frac{\hat{q}}{\hat{1}}\Phi_{\bar{2}}\delta\left\{D(-\ 345)\Phi_{\bar{3}}\Phi_{\bar{4}}\Phi_{\bar{5}}\right\}D(-\ 12)+\frac{\hat{q}}{\hat{1}}\delta\left\{D(-\ 23)\Phi_{\bar{2}}\Phi_{\bar{3}}\right\}D(-\ 45)\Phi_{\bar{4}}\Phi_{\bar{5}}\\
&+\frac{\hat{q}}{\hat{1}}\Phi_{\bar{2}}\delta\left\{D(-\ 34)\Phi_{\bar{3}}\Phi_{\bar{4}}\right\}D(-\ 51)\Phi_{\bar{5}}+\frac{\hat{q}}{\hat{1}}\Phi_{\bar{2}}\Phi_{\bar{3}}\delta\left\{D(-\ 45)\Phi_{\bar{4}}\Phi_{\bar{5}}\right\}D(-\ 12)\\
&+\frac{\hat{q}}{\hat{1}}\delta\Phi_{\bar{2}}D(-\ 3451)\Phi_{\bar{3}}\Phi_{\bar{4}}\Phi_{\bar{5}}+\frac{\hat{q}}{\hat{1}}\Phi_{\bar{2}}\delta\Phi_{\bar{3}}D(-\ 4512)\Phi_{\bar{4}}\Phi_{\bar{5}}\\
&+\frac{\hat{q}}{\hat{1}}\Phi_{\bar{2}}\Phi_{\bar{3}}\delta\Phi_{\bar{4}}D(-\ 5123)\Phi_{\bar{5}}+\frac{\hat{q}}{\hat{1}}D(-\ 1234)\Phi_{\bar{2}}\Phi_{\bar{3}}\Phi_{\bar{4}}\delta\Phi_{\bar{5}}\bigg\}+\cdots
\end{split}\end{equation} 
It is not hard to envisage that this continues to all orders and we can therefore hypothesise a final result to all orders in perturbation theory, given by (\ref{eq:DeltaPhi})

\bibliography{references} 
\bibliographystyle{jhep}
\end{document}